\documentclass[]{aa}  
\usepackage{txfonts}
\usepackage[utf8]{inputenc}
\usepackage{longtable}

\usepackage[version=3]{mhchem}
\usepackage{threeparttable, tablefootnote}
\usepackage{xcolor}
\usepackage{multicol}
\usepackage{graphicx}
\usepackage{txfonts}
\usepackage{amssymb}
\usepackage{rotating}
\usepackage{gensymb}

\usepackage[colorlinks = true,
            linkcolor = blue,
            urlcolor  = blue,
            citecolor = blue,
            anchorcolor = blue]{hyperref}
            
\begin{document} 

\title{Tracing snowlines and C/O ratio in a planet-hosting disk}
\subtitle{ALMA molecular line observations towards the HD~169142 disk}

\author{Alice S. Booth \inst{1} \and
Charles J. Law\inst{2} \and
Milou Temmink \inst{1} \and
Margot Leemker \inst{1} \and
Enrique Mac\'ias \inst{3}}

\institute{Leiden Observatory, Leiden University, 2300 RA Leiden, the Netherlands \and
Center for Astrophysics - Harvard \& Smithsonian, 60 Garden St., Cambridge, MA 02138, USA  \and
European Southern Observatory, Karl-Schwarzschild-Strasse 2, 85748 Garching bei M{\"u}nchen, Germany \\
\email{abooth@strw.leidenuniv.nl; alice.booth@cfa.harvard.edu} 
}

\titlerunning{Tracing snowlines and C/O ratio in a planet-hosting disk}
\authorrunning{A. S. Booth et al.}

\abstract{
The composition of a forming planet is set by the material it accretes from its parent protoplanetary disk. Therefore, it is crucial to map the chemical make-up of the gas in disks to understand the chemical environment of planet formation. This paper presents molecular line observations taken with the Atacama Large Millimeter/submillimeter Array of the planet-hosting disk around the young star HD~169142. We detect \ce{N_2H^+}, \ce{CH_3OH}, [CI], \ce{DCN}, \ce{CS}, \ce{C^{34}S}, \ce{^{13}CS}, \ce{H_2CS}, \ce{H_2CO}, \ce{HC_3N} and \ce{c-C_3H_2} in this system for the first time. Combining these data with the recent detection of SO and previously published \ce{DCO^+} data, we estimate the location of \ce{H_2O} and \ce{CO} snowlines and investigate radial variations in the gas phase C/O ratio. We find that the HD~169142 disk has a relatively low \ce{N_2H^+} flux compared to the disks around Herbig stars HD~163296 and MWC~480 indicating less CO freeze-out and place the CO snowline beyond the millimetre disk at $\approx150$~au. The detection of \ce{CH_3OH} from the inner disk is consistent with the \ce{H_2O} snowline being located at the edge of the central dust cavity at $\approx20$~au. The radially varying CS/SO ratio across the proposed \ce{H_2O} snowline location is consistent with this interpretation. Additionally, the detection of \ce{CH_3OH} in such a warm disk adds to the growing evidence supporting the inheritance of complex ices in disks from the earlier, colder stages of star formation. Finally, we propose that the giant HD~169142~b located at 37~au is forming between the \ce{CO_2} and \ce{H_2O} snowlines where the local elemental make of the gas is expected to have C/O$\approx$1.0.}
   \maketitle
%
\section{Introduction}

The elemental make-up of an exoplanet's atmosphere is set by the material it accretes locally from its parent protoplanetary disk \citep{2021PhR...893....1O}.  With ground-based high-resolution spectroscopy and the James Webb Space Telescope (JWST), we can infer the C/H, O/H and C/O ratios in the atmospheres of giant exoplanets \citep[e.g.][]{2020A&A...633A.110G, 2020AJ....160..150W, 2022AJ....163..159R, 2023arXiv230314206B}. These observations now provide a direct link to unravelling the formation history of planets \citep[e.g. see][for a review]{2023ESC.....7..260E}. This is particularly important, as many of the gas-giant planets detected may not have formed at their current location and have undergone migration \citep{2018ARA&A..56..175D}. 
A key piece of this puzzle is to measure the elemental make-up of disks and in particular locating snowlines (\ce{CO} and \ce{H_2O}) at the time of planet formation.  

The local gas and ice composition in a protoplanetary disk is regulated by various physical and chemical processes. The elemental make-up of the disk gas and ice varies as a function of radius and height due to the freeze-out of different molecules \citep{2011ApJ...743L..16O}. The growth, settling, and inward drift of dust transports volatile elements leading to an elevated gas-phase C/O$>$1 in the outer regions of disks and enhanced gas-phase abundances of CO and \ce{H_2O} within their respective snowlines \citep{2015ApJ...807L..32D, 2018ApJ...864...78K, 2019ApJ...883...98Z, 2019MNRAS.487.3998B, 2020ApJ...903..124B}. Directly measuring the locations of snowlines is non-trivial. The most abundant and easily observable isotopologues of CO tend to be optically thick and trace gas in a layer above the disk midplane. The \ce{H_2O} emission lines accessible at sub-millimetre wavelengths are weak and difficult to detect in Class II disks \citep{2019ApJ...875...96N, 2021ApJ...918L..10B}. 

Luckily we can use other molecules as chemical tracers of both the CO and \ce{H_2O} snowlines. The cations \ce{DCO^+} and \ce{N_2H^+} have been used as tracers of the CO snowline and of the two species \ce{N_2H^+} is the more robust tracer \citep{2013Sci...341..630Q, 2015ApJ...813..128Q}. \ce{N_2H^+} is destroyed by gas-phase \ce{CO} and is thus abundant in the outer disk regions where CO is frozen-out \citep{2017A&A...599A.101V}. Due to a similar chemical relationship, \ce{HCO^+} is chemical tracer of the \ce{H_2O} snowline \citep[e.g.,][]{2013ApJ...779L..22J,2021A&A...646A...3L}. Additionally, the presence of methanol (\ce{CH_3OH}) indicates warm enough temperatures for the sublimation of \ce{H_2O} \citep{2021NatAs...5..684B, 2021A&A...651L...5V}. To complement these chemical tracers, the variations in C/O across disks due to snowlines can be traced by observing pairs of molecules, e.g, SO and CS or NO and CN \citep{2019ApJ...886...86L, 2021A&A...651L...6B, 2021ApJS..257...12L, 2023arXiv230300768L}, or by detecting molecules that have enhanced formation in elevated C/O environments, e.g., \ce{C_2H} and \ce{c-C_3H_2} \citep{2019A&A...631A..69M, 2021ApJS..257....6G, 2021ApJS..257....7B, 2023NatAs...7...49C}.

HD~169142 is a nearby (114.0~pc) F1 star hosting an almost face on (inclination of 13$^{\circ}$, position angle of 5$^{\circ}$) protoplanetary disk \citep{2006AJ....131.2290R,2017A&A...600A..72F, 2018A&A...616A...1G, 2018A&A...620A.128V}. Multi-wavelength dust observations have revealed multiple dust rings in this transition disk that are likely carved by forming giant planets \citep{2017A&A...600A..72F, 2017ApJ...850...52P, 2017ApJ...838...97M, 2019ApJ...881..159M, 2019AJ....158...15P}. CO gas detected within the central dust cavity points to the presence of a giant planet or low-mass companion within 10~au \citep{2018tcl..confE..64C, 2022A&A...663A..23L, 2022MNRAS.517.5942G, 2022MNRAS.510..205P}. In the outer disk, in a dust and gas gap, there is evidence supporting a giant planet located at 37~au \citep{2019A&A...623A.140G, 2022MNRAS.517.5942G, 2023arXiv230211302H}. Studies of the kinematics of the CO gas in the outer disk show strong deviations from Keplerian motion in the vicinity of the planet \citep{2022MNRAS.517.5942G} and recently, \citet{Law2023} show evidence for shocks associated with this planet traced in SiS and SO. In this paper, we present Atacama Large Millimeter/submillimeter Array (ALMA) observations of the HD~169142 disk where we detect 13 different species. Using these data, we infer the elemental make-up of the gas and location of both the CO and \ce{H_2O} snowlines in relation to these forming planets.

\section{Observations}

We have compiled ALMA observations towards the HD~169142 disk over Bands 6, 7 and 8. Full details, including, the observation dates, number of executions, baselines, integration times and maximum recoverable scales are listed in Table A.1. 
All data sets were first calibrated by the ALMA staff and the subsequent data reduction, representative angular resolution and the molecules detected in each data set are summarised below: 

\textit{Project 2012.1.00799.S} (PI: Honda, M.) are Band 7 observations at $\approx$0.2" and in these data, we detect the \ce{CH_3OH} $J=11_{(1,10)}-11_{(0,11)}$ and the SO $J=8_8-7_7$ where the SO was first presented in \citet{Law2023} along with CO isotopologues (\ce{^{12}CO} and \ce{^{13}CO} $J=3-2$). The continuum data from this program were initially presented in \citet{2019ApJ...881..159M}. For further details about the data reduction, please refer to \citet{Law2023}.

\textit{Project 2015.1.00806.S} (PI: Pineda, J.) are Band 7 observations at $\approx$0.03" and in these data, we detect \ce{CS} $J=6-5$. These observations also provide long baselines of the CO isotopologues (\ce{^{12}CO} and \ce{^{13}CO} $J=3-2$). For further details on the data reduction, please refer to \citet{Law2023}.

\textit{Project 2016.1.00344.S} (PI: Perez, S.) are Band 6 observations at $\approx$0.2". This project also has higher angular resolution data ($\approx$0.02") associated with it \citep[see][for the continuum observations]{2019AJ....158...15P} but as we are searching for weak lines emission we only make use of the shorter baseline data, which are optimal for our science goals.
These observations consist of 5 spectral windows with 3 dedicated to the CO lines (\ce{^{12}CO}, \ce{^{13}CO}, \ce{C^{18}O} $J=2-1$) and two continuum windows centred at 218.004~GHz and 232.004~GHz. We obtained the calibrated measurement sets from the ALMA archive and performed one round of phase-only self-calibration (interval 60 seconds) on the data after flagging the line containing channels. This resulted in only a 20\% improvement in the continuum signal-to-nose ratio. In these data, we detect lines of \ce{H_2CO} $J=3_{(0,3)}-2_{(0,2)}$ and $J=3_{(2,2)}-3_{(2,1)}$, DCN $J=3-2$, \ce{HC_3N} $J=24-23$ and \ce{c-C_3H_2} $J=6_{(0,6)}-5_{(1,5)}$ and $J=6_{(1,6)}-5_{(0,5)}$ in the 218.004~GHz continuum spectral window (see Figure~C1). 

\textit{Project 2016.1.00346.S} (PI: Tsukagoshi, T.) are Band 8 observations at $\approx$1.0" and we detect \ce{CS} $J=10-9$ and [CI] $^{3}P_1-^{3}P_0$. The dataset contains four spectral windows, which are centred at 478.651 GHz, 480.568 GHz, 489.785 GHz and 492.195 GHz with the latter two the line spectral windows at $\sim0.15$ km s$^{-1}$ spectral resolution. We obtained the calibrated measurement set from the ALMA archive and then performed self-calibration on the data. This consisted of three rounds of phase-only calibration using solution lengths (60, 20 and 6 seconds). The self-calibration yielded a signal-to-noise improvement of 40\% for the continuum.

\textit{Project 2018.1.01237} (PI: Mac\'ias, E.) are Band 7 observations at $\approx$0.8" resolution. These data target and detect \ce{DCO+} $J=4-3$ and \ce{N_2H+} $J=3-2$ and also detect
\ce{DCN} $J=4-3$, \ce{^{13}CS} $J=6-5$, \ce{C^{34}S} $J=6-5$, \ce{H_2CS} $J=8_{(1,7)}-7_{(1,6)}$, \ce{H_2CO} $J=4_{(0,4)}-3_{(0,3)}$ and,
 \ce{CH_3OH} $J=9_{(-1,9)}-8_{(0,8)} $ and $J=6-5$ blend of lines. 
Four spectral windows were used, centred at 279.515 GHz (resolution of $\sim0.15$ km s$^{-1}$), 277.990 GHz ($\sim1.2$ km s$^{-1}$), 288.147 GHz ($\sim0.29$ km s$^{-1}$) and 289.990 GHz ($\sim17$ km s$^{-1}$). 
The data were then phase-only self-calibrated using the continuum emission. For this, the line emission channels were first flagged to create a continuum-only dataset. 
Three rounds of phase-only self-calibration were applied, with solution intervals going down from \textit{inf} (i.e., the scan duration) to 30 s. Overall, the SNR of the continuum improved from $\sim900$ to $\sim1600$. The solutions were then applied to the unflagged data.
A weak detection of \ce{N_2H^+} in the images was confirmed via both matched filtering and spectral stacking with \textit{GoFish} \citep[see Figure B1; ][]{2018AJ....155..182L,GoFish}. The \ce{DCN}, \ce{C^{34}S}, \ce{H_2CO} and one of the \ce{CH_3OH} lines are observed in a continuum spectral window at $\approx$17~km~s$^{-1}$ spectral resolution (see Figure C1).

The continuum was subtracted using \textit{uvcontsub} with a fit order of 1 after excluding the line containing channels. The line containing channels of the continuum spectral windows were identified in an iterative process. First, the continuum was subtracted assuming no lines were present, we made a dirty image of the line, extracted spectra, identified any lines and then made a new continuum subtracted measurement set flagging these channels. We performed all of the imaging in CASA version 6.2 \citep{2007ASPC..376..127M} and used Keplerian masks for the tCLEAN mask (generates with the code from \citealt{rich_teague_2020_4321137}) for lines at a spectral resolution $<$2~km~s$^{-1}$ and hand-drawn elliptical tCLEAN masks for other lines that were observed at a coarser spectral resolution as total line width is only a few km~s$^{-1}$ across the disk. Table 1 lists all of the properties of the molecular lines detected and the resulting images. All lines in the 2016.1.00344.S data aside from \ce{HC_3N} were tapered to a 0.4" beam to increase signal-to-noise. 

\begin{table*}[h!]
\scriptsize
\caption{Molecular transitions and image properties for the detected lines in the HD~169142 disk.}
\centering
\begin{tabular}{c c c c c c c c c c c}
        \hline \hline
Species  & Transition & Frequency & E$_{\mathrm{up}}$ &  E$_{\mathrm{Aul}}$ &  g$_u$ & \texttt{robust} &Beam & $\delta$v & rms & Int. Flux\\ 
&  &  (GHz) & (K) & (s$^{-1}$) &  & &   & (km s$^{-1}$)  & (mJy beam$^{-1}$) & (mJy km s${^{-1}}$) \\ \hline 

\ce{[C I]} & $^{3}P_1-^3P_0$            & 492.160651   & 24    & 7.994$\times10^{-4}$ & 3  & 2.0& 1.07"$\times$0.51" (81\degree) & 0.15 & 71.9 & (7.4$\pm$0.7)$\times10^{3}$ \\

\ce{DCO^+}& $J=4-3$                     & 288.143858   & 35    & 5.858$\times10^{-3}$ & 9  & 0.5 & 0.72"$\times$0.46" (89\degree) & 0.26  & 2.64 & 970$\pm$100  \\    

\ce{DCN}& $J=3-2$                       & 217.238539   & 21    & 4.575$\times10^{-4}$ & 21 & 2.0, taper & 0.41"$\times$0.37" (86\degree)& 2.7  & 0.67 & 93.0.0$\pm$10  \\    
\ce{DCN}& $J=4-3$                       & 289.644917   & 35    & 1.124$\times10^{-3}$ & 27 & 0.5 & 0.73"$\times$0.46" (89\degree)    & 16.2  & 0.24 & 240.0$\pm$20  \\  

\ce{N_2H+}&  $J=3-2$                    & 279.511701   & 27    & 1.259$\times10^{-3}$ & 63 & 2.0 & 0.91"$\times$0.58" (85\degree)  & 0.14 & 2.95 & 55.0$\pm$6 \\

\ce{CS}$\dagger$ & $J=7-6$              & 342.882850   & 66    & 8.368$\times10^{-4}$ & 15 & 0.5 & 0.29"$\times$0.26"(37\degree) & 0.85 & 11.7 & ${\gtrsim}$7$\times10^{3}$ \\
\ce{CS} & $J=10-9$                      & 489.750921   & 129   & 2.489$\times10^{-3}$ & 21 & 0.5 & 0.94"x0.40" (83\degree) & 0.15 & 130.7 & (4.3$\pm$0.4)$\times10^{3}$ \\
\ce{C^{34}S}& $J=6-5$                   & 289.209066   & 38    & 4.812$\times10^{-4}$ & 13 & 0.5 & 0.73"$\times$0.46" (89\degree)    & 16.2   & 0.23 &   261.0$\pm$30    \\
\ce{^{13}CS}& $J=6-5$                   & 277.455405   & 47    & 4.399$\times10^{-4}$ & 26 & 0.5 & 0.73"$\times$0.46" (89\degree)     & 1.10   & 1.1 &   51$\pm$5  \\

\ce{SO}& $J=8_8-7_7$                    & 344.310612   & 88    & 5.186$\times10^{-4}$ & 17 & 2.0  & 0.19"$\times$0.14" (84)    & 0.46 & 1.50 &  120$\pm$10    \\ 

\ce{H_2CS}& $J=8_{(1,7)}-7_{1,6}$       & 278.887661   & 73    & 3.181$\times10^{-4}$ & 51 & 0.5 & 0.73"$\times$0.46" (89\degree)  & 1.10   & 1.07 & 140$\pm$10 \\ 

\ce{H_2CO}& $J=3_{0,3}-2_{0,2}$     & 218.222192   & 21    & 2.818$\times10^{-4}$ & 7  & 2.0, taper & 0.41"$\times$0.37" (86\degree)  & 2.7   & 0.52 &  480$\pm$50     \\
\ce{H_2CO}& $J=3_{2,2}-3_{2,1}$     & 218.475632   & 68    & 1.571$\times10^{-4}$ & 7  & 2.0, taper & 0.41"$\times$0.37" (86\degree)  & 2.7  & 0.51  &   55$\pm$6     \\
\ce{H_2CO}& $J=4_{0,4}-3_{0,3}$     & 290.623405   & 35    & 6.902$\times10^{-4}$ & 9  & 0.5 & 0.73"$\times$0.46" (89\degree)    & 16.2   & 0.26&  900$\pm$90    \\

\ce{CH_3OH} & $J=4_{2,2}-3_{1,2}$ & 218.440050           & 45   & 4.687$\times10^{-5}$ & 36 & 2.0 & 0.20"$\times$0.15" (88)    & 0.46 & 1.44 &  65$\pm$7\\ 
\ce{CH_3OH}& $J=9_{(-1,9)}-8_{(0,8)}$   & 278.304510   & 110   & 7.685$\times10^{-5}$ & 76 & 0.5 &  0.73"$\times$0.46" (89\degree)       & 1.10   & 1.05 &  13$\pm$1    \\
\ce{CH_3OH}& blend$^*$                  & 290.209700   & 112   & 7.994$\times10^{-5}$ & 52 & 0.5 & 0.73"$\times$0.46" (89\degree)    & 16.2   & 0.24 &   104$\pm$10       \\
\ce{CH_3OH}& $J=11_{(1,10)}-11_{(0,11)}$& 331.502370   & 169   & 3.929$\times10^{-4}$ & 92 & 2.0 & 0.20"$\times$0.15" (88)    & 0.46 & 1.44 & 44$\pm$10  \\

\ce{HC_3N} &  $J=24-23$                 & 218.324723   & 131   & 8.261$\times10^{-4}$ & 49 & 2.0 & 0.26"$\times$0.21" (74\degree)& 2.7  & 0.45 &    171$\pm$20   \\

\ce{c-C_3H_2} & $J=6_{(0,6)}-5_{(1,5)}$ & 217.822148   & 39    & 5.934$\times10^{-4}$ & 13 & 2.0, taper & 0.41"$\times$0.37" (86\degree)& 2.7  & 0.59 &   59$\pm$6$\dagger$$\dagger$ \\
\ce{c-C_3H_2} & $J=6_{(1,6)}-5_{(0,5)}$ & 217.822148   & 39    & 5.934$\times10^{-4}$ & 39 & 2.0, taper & 0.41"$\times$0.37" (86\degree)& 2.7  & 0.59 &   59$\pm$6$\dagger$$\dagger$ \\

\hline 
\end{tabular}        
 \begin{tablenotes}
\item{
All of the molecular data are taken from the Cologne Database for Molecular Spectroscopy \citep[CDMS,][]{2005JMoSt.742..215M}. Since we are comparing molecular lines from a variety of different sets of observations we take the errors on the integrated fluxes to be the 10\% absolute flux calibration uncertainty of ALMA. \newline
For reference, the \ce{C^{18}O} $J=2-1$ disk integrated flux from \citet{2017A&A...600A..72F} is 3.9$\pm$0.5 Jy~km s$^{-1}$.
$^*$ Blended \ce{CH_3OH} emission from up to 6 lines where the  E$_{\mathrm{up}}$, E$_{\mathrm{Aul}}$, and g$_u$ listed are of the strongest line $J=6_{(-3,4)}-5_{(-3,3)}$. Other lines are: $J=6_{(3,3)}-5_{(3,2)}$, $J=6_{(3,4)}-5_{(3,3)}$, $J=6_{(4,2)}-5_{(4,1)}$, $J=6_{(2,5)}-5_{(2,4)}$ and $J=6_{(3,3)}-5_{(3,2)}$.  \newline
$\dagger$ The CS $J=7-6$ image was generated with a Gaussian \textit{uv}-taper of 0.25" and then further smoothed in the image plane with the CASA \texttt{imsmooth} task using a Gaussian kernel of 0.2". Due to 
the lack of short-baseline data we consider this flux a lower limit. \newline
$\dagger \dagger$ flux is of two blended lines. \newline}
\end{tablenotes}
 \label{tab:1}
\end{table*}

\begin{figure*}
    \centering
    \includegraphics[trim=2cm 4.5cm 2cm 5cm, clip,width=0.95\hsize]{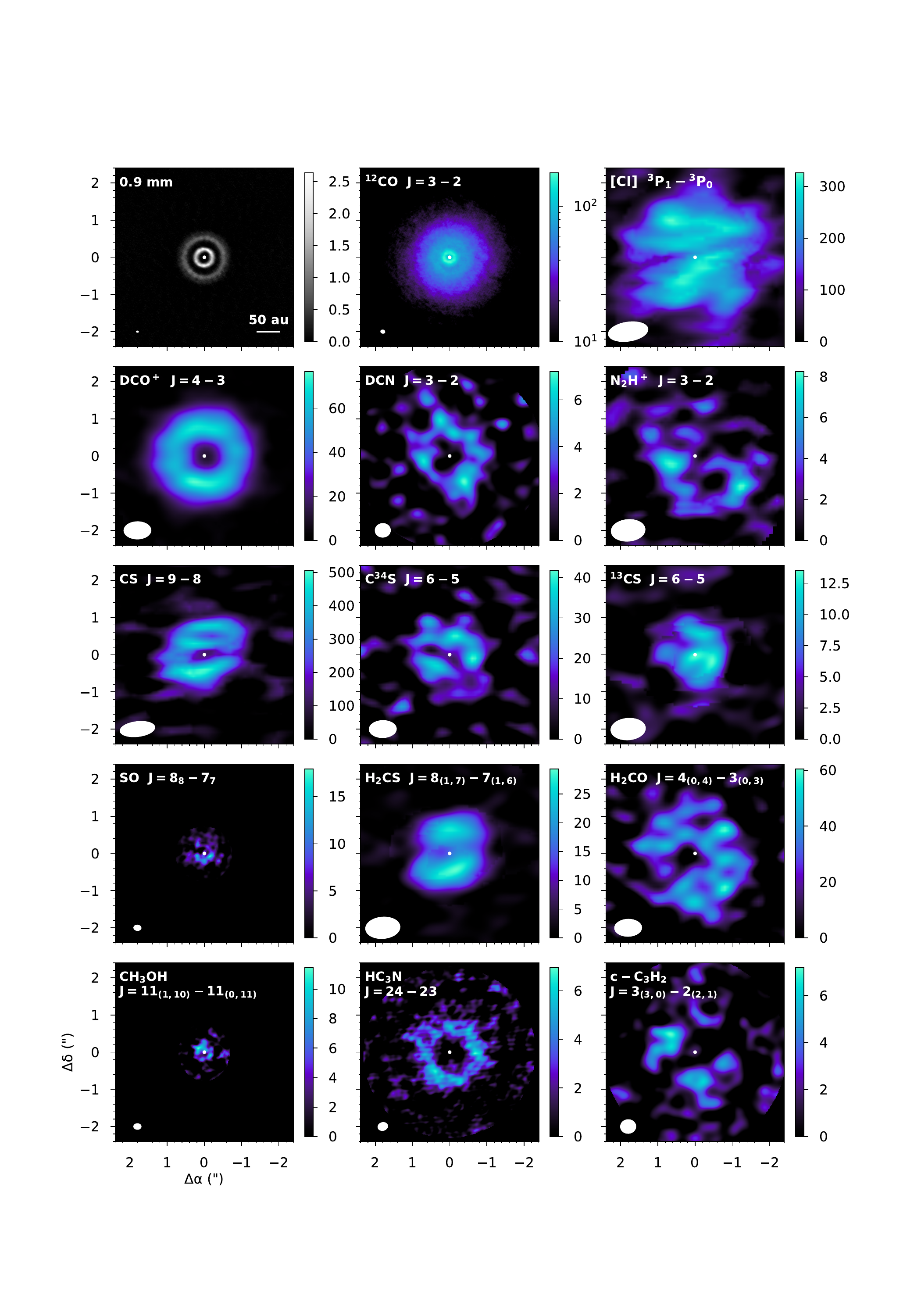}
    \caption{Integrated intensity maps of the millimetre dust and molecular line emission from the HD~169142 protoplanetary disk. Units of the colour bars are $\mathrm{mJy~beam^{-1}}$ for the continuum image and $\mathrm{mJy~beam^{-1}~km~s^{-1}}$ for the line images aside from \ce{^{12}CO} which is $\mathrm{Jy~beam^{-1}~km~s^{-1}}$ . The beam of each of the observations is shown in the bottom left-hand corner of the plots and the central dot denotes the position of the star.}
    \label{fig:my_label}
\end{figure*}

\begin{figure*}
    \centering
    \includegraphics[trim=0cm 0cm 0cm 0cm, clip,width=0.93\hsize]{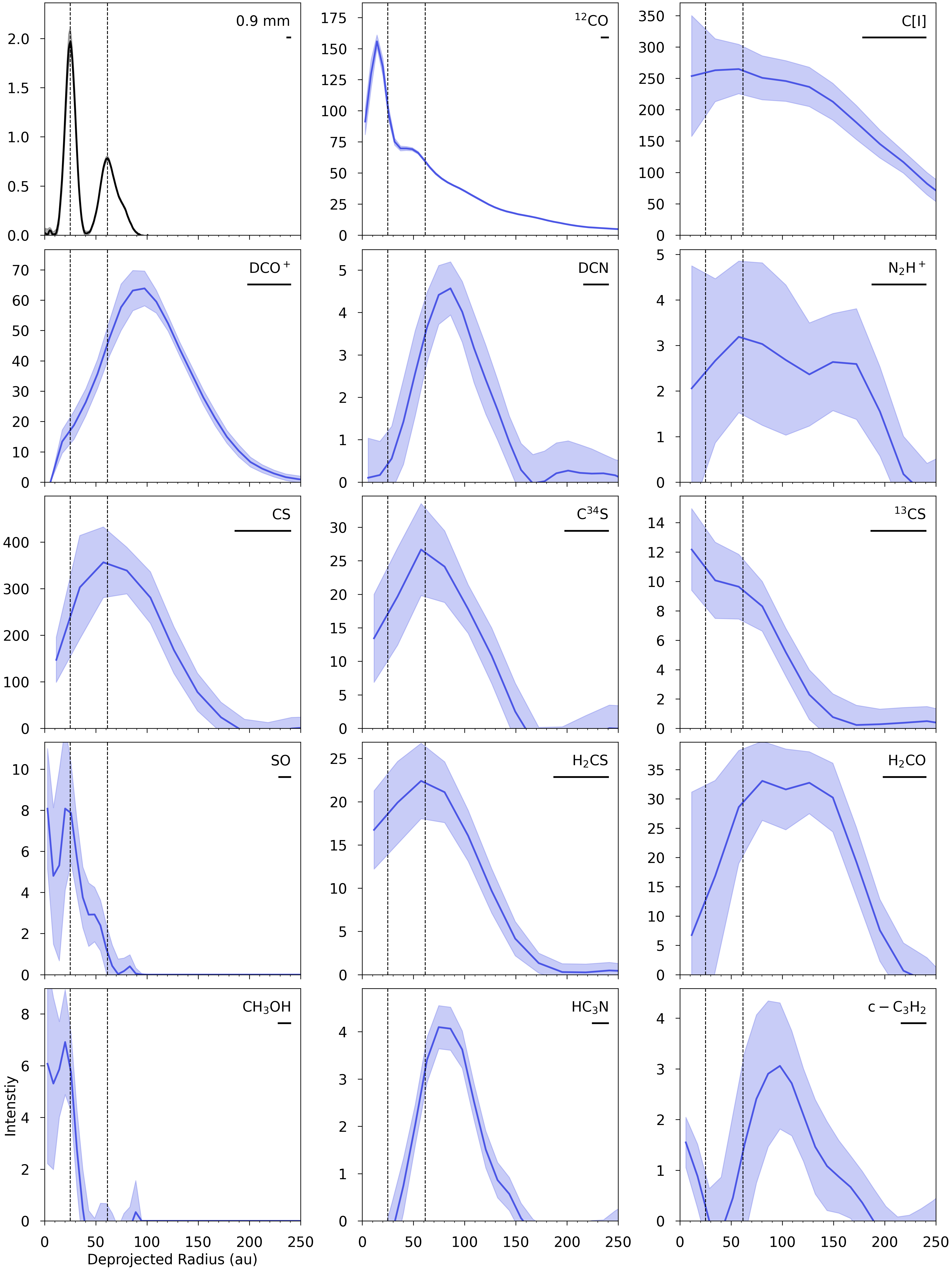}
    \caption{Azimuthally averaged radial intensity profiles for the 0.9mm continuum emission and molecular lines from the HD169142 disk. The specific transitions are the same as in Figure~1. Vertical lines mark the peak locations of the two bright continuum rings. The bar in the top right of the images is the 0.5$\times$ the major axis of the beam as listed in Table~1. Units of the y-axis are 
    mJy~beam$^{-1}$ for the continuum, Jy~beam$^{-1}$~km~s$^{-1}$ for the \ce{^{12}CO} and mJy~beam$^{-1}$~km~s$^{-1}$ for the rest of the lines.}
    \label{fig:my_label}
\end{figure*}

\section{Results}

All integrated intensity maps were generated with the same Keplerian or elliptical masks used in the imaging with no flux clipping to ensure accurate fluxes. Figure~1 presents the resulting maps for the different species detected in the HD~169142 disk. Where multiple transitions are detected, we show the "best image" and all of the detected transitions are shown in Figure D1.  For comparison, we also include the 0.9~mm continuum image and \ce{^{12}CO} map from \citet{Law2023}. All of the newly reported molecules show varying ring-like morphologies with the SO and \ce{CH_3OH} being the most compact. Figure~2 shows the azimuthally-averaged radial profiles for the same lines as presented in Figure~1. These profiles highlight a clear anti-correlation between SO and CS.

Table~\ref{tab:1} lists the disk-integrated fluxes for all of the detected lines. 
We take the errors on these fluxes to be $\pm$10\% due to the uncertainty on the absolute flux calibration of ALMA. We estimate disk-averaged column densities for all the molecules detected using these disk-integrated fluxes following the now standard methods \citep[e.g.,][]{2018ApJ...859..131L}.
This assumes the gas is in local thermodynamic equilibrium and that the line emission is optically thin.
For most species, we only detect one transition therefore we calculate all of the column densities at 30 and 60~K for the molecules we expect to be tracing cooler gas and 100~K for the SO and \ce{CH_3OH} which are emitting from the warmer inner disk. Where multiple lines are detected we take an average of the calculated column density. 
We exclude the \ce{H_2CO} $J=3_{(2,2)}-3_{(2,1)}$ and \ce{CH_3OH}  $J=4_{(2,2)}-_{(1,2)}$ lines from this analysis as they are low signal-to-noise detections in the spectra (see Figure C1). Additionally, the \ce{H_2CO} $J=3_{(2,2)}$-$3_{(2,1)}$ line does not share the same emission morphology as the other two strongly detected \ce{H_2CO} lines (see Figure~D1). 

We also compute radial column densities from the emission profiles presented in Figure~2. This is done for SO, CS (using the optically thinner \ce{C^{34}S} and assuming \ce{^{32}S/\ce{^{34}S}} = 22, e.g.; \citealt{1999RPPh...62..143W}), \ce{H_2CO} and \ce{CH_3OH} as we aim to investigate any radial variations in N(CS)/N(SO) and N(\ce{CH_3OH})/N(\ce{H_2CO}) ratios. Since we do not have any excitation information we assume a fixed temperature of 60~K. The SO and \ce{CH_3OH} are likely warmer than this and the CS and \ce{H_2CO} may be cooler but given the data in hand this intermediate temperature is the simplest assumption to make. Figure~3 shows the radial column densities at the native angular resolution for each of the four molecules alongside a version where the \ce{SO} and \ce{CH_3OH} images are convolved to the same angular resolution of the \ce{^{34}CS} and \ce{H_2CO} data using CASA tool \textit{imsmooth}. We find that the N(CS)/N(SO) ratio is increasing as a function of radius and the N(\ce{CH_3OH})/N(\ce{H_2CO}) ratio is decreasing as a function of radius (see Figure~3). These ratios in the context of the disk chemistry and forming planets in the HD~169142 disk will be discussed in Section 4. 



\begin{table}
\small
    \centering
    \caption{Disk-averaged column densities assuming a circular emitting area of 2" for all species aside from \ce{CH_3OH} and SO where we use 0.3". A range of T$\mathrm{_{ex}}$ are explored for all species and from DCN, \ce{DCO^+}, \ce{H_2CO} and \ce{CH_3OH}.
    Errors in column densities are propagated from the assumed 10\% uncertainty in the disk-integrated fluxes.}
    \begin{tabular}{c c c c}
    \hline \hline
        Species &  & N (cm$^{-2}$) &  \\
         & $\mathrm{T_{ex}:}$ 30~K & 60~K  & 100~K  \\ \hline
        \ce{[C I]} & (5.1$\pm$0.5)$\times10^{16}$ &  (5.0$\pm$0.5)$\times10^{16}$ & - \\
         \ce{DCO^+}&  (3.0$\pm$0.3)$\times10^{11}$ &  (3.7$\pm$0.4)$\times10^{11}$ & - \\
         \ce{DCN}&  (2.6$\pm$0.3)$\times10^{11}$ &  (3.1$\pm$0.3)$\times10^{11}$ & - \\
         \ce{N_2H^+} & (4.5$\pm$0.5)$\times10^{10}$ & (5.7$\pm$0.6)$\times10^{10}$  & - \\
        \ce{CS} & (3.8$\pm$0.4)$\times10^{13}$ & (8.2$\pm$0.8)$\times10^{12}$ & - 
        \\
         \ce{C^{34}S }  & (8.5$\pm$0.9)$\times10^{11}$ & (8.9$\pm$0.9)$\times10^{11}$  & - \\
         \ce{^{13}CS }  & (2.5$\pm$0.3)$\times10^{11}$ & (2.3$\pm$0.3)$\times10^{11}$  & - \\
            \ce{SO} &  - & - & (2.1$\pm$0.3)$\times10^{13*}$ \\ 
         \ce{H_2CS}&   (4.2$\pm$0.4)$\times10^{12}$ & (3.5$\pm$0.4)$\times10^{12}$  & - \\
         \ce{H_2CO}&  (1.0$\pm$0.1)$\times10^{13}$ & (1.6$\pm$0.2)$\times10^{13}$  & - \\
            \ce{CH_3OH} &  - & - & (3.3$\pm$0.3)$\times10^{14}$ \\ 
            \ce{HC_3N}   & (1.0$\pm$0.1)$\times10^{13}$ & (2.2$\pm$0.2)$\times10^{12}$  & - \\
         \ce{c-C_3H_2}   & (5.6$\pm$0.6)$\times10^{11}$ & (8.4$\pm$0.8)$\times10^{11}$  & - \\ \hline
         \hline
    \end{tabular}
         \begin{tablenotes}
         \item{
    $^{*}$ taken from \citet{Law2023}. \\}
            \end{tablenotes}
    \label{tab:my_label}
\end{table}

\begin{figure*}[h!]
    \centering
    \includegraphics[width=0.95\hsize]{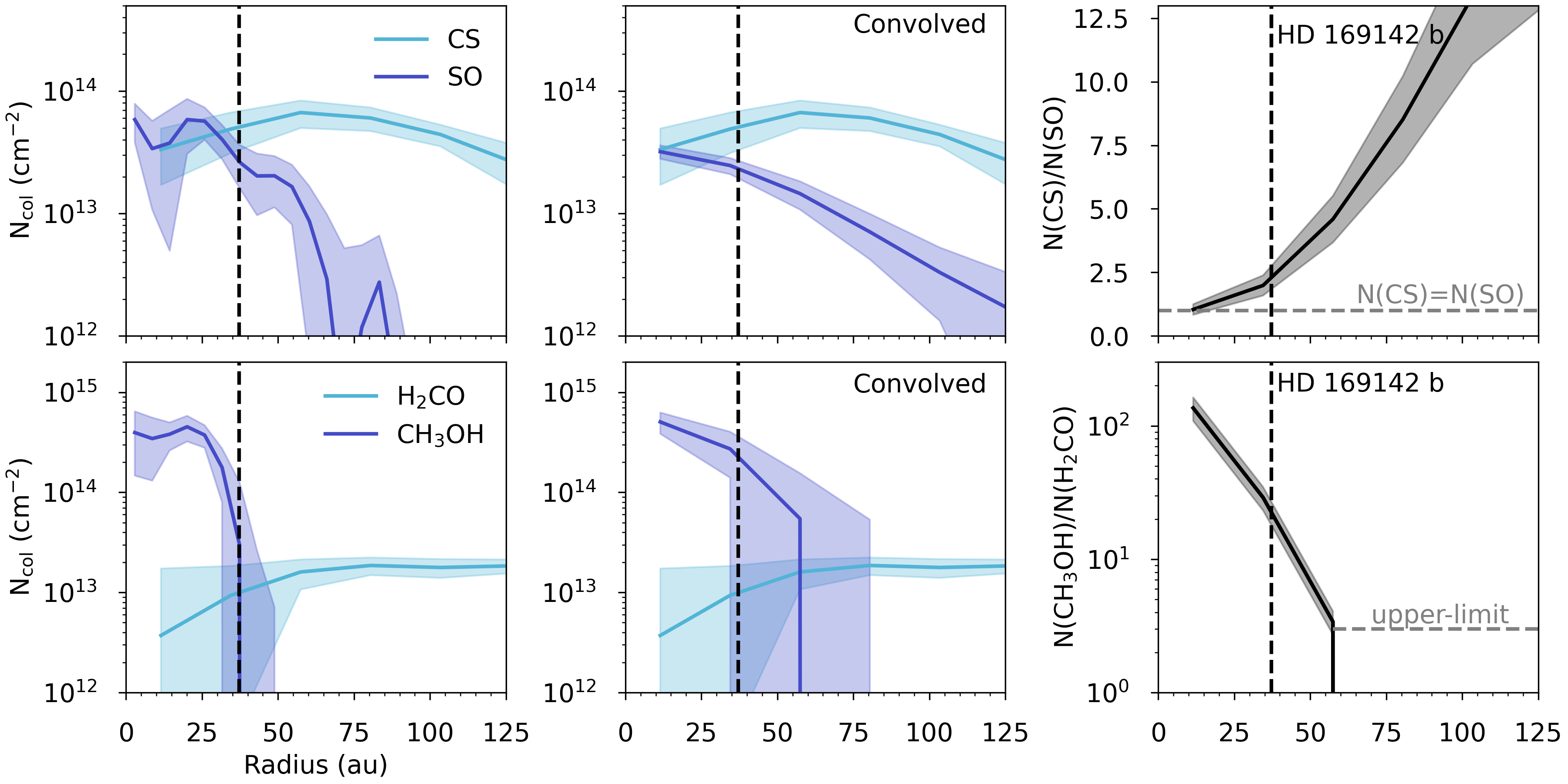}
    \caption{Radial profiles of column densities and column density ratios for SO, CS, \ce{CH_3OH} and \ce{H_2CO} in the HD~169142 disk. These are computed from the radial emission profiles in Figure~2 assuming a temperature of 60~K. Left: at the native resolution of the data, middle: convolved to the same beam, and right: ratios of the convolved profiles. For SO and CS ratio we show where N(CS)/N(SO)=1 with a horizontal line and for \ce{H_2CO} and \ce{CH_3OH} we show the upper-limit on the column density ratio where \ce{CH_3OH} is not detected. The vertical dashed line highlights the radial location of the giant planet HD~169142~b.}
    \label{fig:my_label}
\end{figure*}

\begin{figure*}
    \centering
    \includegraphics[width=0.93\hsize]{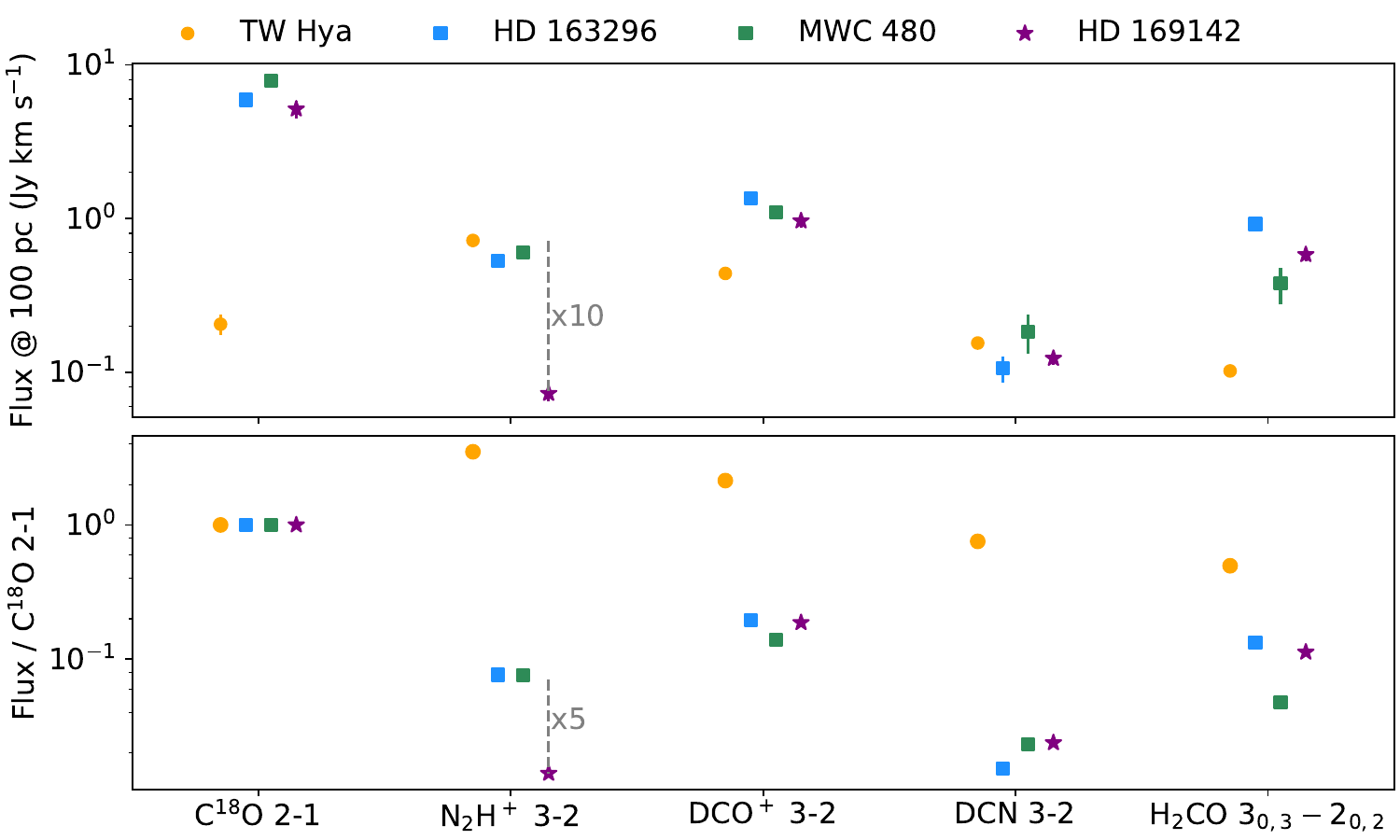}
    \caption{Comparisons of line fluxes and line flux ratios for the TW Hya, HD~163296, MWC~480 and HD~169142 disks. Data are compiled from this work, \citet{2015ApJ...813..128Q, 2018A&A...614A.106C, 2020ApJ...890..142P, 2020ApJ...893..101L, 2021ApJ...908....8C, 2021ApJS..257....1O, 2023ApJ...943...35M, 2023arXiv230302167P}. In the top panel line fluxes and their associated errors are scaled to 100~pc for each disk taking TW~Hya at 60~pc, HD~169142 at 114~pc, MWC~480 at 160~pc, and HD~163296 at 101~pc \citep{2018A&A...616A...1G, 2018A&A...620A.128V}. }
    \label{fig:4}
\end{figure*}

\section{Discussion}

We have detected a range of molecules from the HD~169142 disk that can be used to unravel the physical and chemical conditions in the system. In this section, we discuss the locations of the \ce{CO} and \ce{H_2O} snowlines, the evidence for radial variations in C/O, and how this connects to the composition of the forming planet(s) in this disk.

\subsection{Locating the CO and \ce{H_2O} snowlines}

The CO snowline in the HD~169142 disk has been previously investigated using observations of \ce{DCO^+} \citep{2017ApJ...838...97M, 2018A&A...614A.106C}. The dust temperature in the fiducial model of this disk from \citet{2018A&A...614A.106C} only reaches the freeze-out temperature of CO ($\approx$20~K) in the midplane of the very outer disk at $\approx$150~au and there is no complete freeze-out of CO indicating a CO snowline. In this work, we present the detection of \ce{N_2H^+} which is a more robust tracer of the CO snowline \citep[e.g.,][]{2015ApJ...813..128Q}. The \ce{N_2H^+} emission extends to $\approx$200~au. \ce{N_2H^+} does not solely trace the CO snowline in disks and can be present in the warm molecular layer of the disk where CO is photo-dissociated and \ce{N_2} is not yet photo-dissociated \citep{2017A&A...599A.101V}.
Comparing to the models presented in \citet{2019ApJ...881..127A} and \citet{2022ApJ...926L...2T} the low \ce{N_2H^+} line flux observed towards HD~169142 relative to \ce{C^{18}O} is consistent with minimal CO depletion, i.e., a gas-phase CO/\ce{H_2} abundance of $\approx10^{-4}$. Therefore, the weak \ce{N_2H^+} emission detected may be tracing the gas-phase chemistry in the upper layers of the disk atmosphere inside the CO snowline. 
Figure 4 compares the \ce{C^{18}O} $J=2-1$, \ce{N_2H^+} $J=3-2$, \ce{DCO^+} $J=3-2$, \ce{DCN} $J=3-2$ and \ce{H_2CO} $J=3_{(0,3)}-2_{(0,2)}$ line fluxes observed towards the HD~169142, TW~Hya, MWC~480, and HD~163296 disks.  Scaled to a distance of 100~pc the \ce{N_2H^+} emission from HD~169142 is $\approx$10$\times$ weaker than in both of the Herbig disks and the T-Tauri disk TW~Hya. Relative to the \ce{C^{18}O} $J=2-1$ line flux the \ce{N_2H^+} is $\approx$5$\times$ weaker than in HD~163296 and MWC~480. The depletion of CO beyond the midplane CO snowline has been measured in both the HD~163296 and MWC~480 disks where \citet{2021ApJS..257....5Z} found a drop in CO abundance of at least a factor of 10 at a radius of 65~au and 100~au for each disk, respectively. In comparison, a similar drop in CO abundance is not required in the HD~169142 disk to reproduce the \ce{DCO^+} data \citep{2018A&A...614A.106C}.
That being said, if there is a CO snowline it is expected to be beyond the mm-dust rings and current known giant planet orbits \citep{2017ApJ...838...97M, 2018A&A...614A.106C}.

We can trace the \ce{H_2O} snowline indirectly under the assumption that the \ce{CH_3OH} is thermally desorbed from the dust grains where $\mathrm{T_{dust}}>100$~K. The temperature of optically thick CO gas in the mm-dust cavity is also $>100$~K, which supports this hypothesis \citep[e.g., ][]{2022A&A...663A..23L}. This is also consistent with the model from \citet{2018A&A...614A.106C} which places the \ce{H_2O} midplane snowline at $\approx$20~au, which is at the edge of the inner dust ring (shown in Figure~E1). 

\subsection{Volatile sulphur and C/O}

In the HD~169142 disk, we have detections of SO, CS, \ce{C^{34}S}, \ce{^{13}CS} and \ce{H_2CS}. Using the ratios between the different \ce{CS} isotopologues we can assess if the main isotopologue is optically thick. The inferred column density ratios of \ce{CS}/\ce{C^{34}S} (44$\pm$10 at 30~K, 10$\pm$5 at 60~K) and \ce{CS}/\ce{^{13}CS} (150$\pm$30 at 30~K, 36$\pm$8 at 60~K) compared to the ISM isotope ratios (\ce{^{32}S/\ce{^{34}S}} = 22, \ce{^{12}C}/\ce{^{13}C}=69, e.g.; \citealt{1999RPPh...62..143W}) indicate optically thick CS if the emitting layer of the gas is warm $\approx$60~K. If the gas is cooler (30~K) then the isotopes appear to be less abundant than in the ISM, but they may emit from different layers of the disk and therefore at different temperatures. Observation of CS $J=6-5$ rather than $J=10-9$ to match the \ce{C^{34}S} and \ce{^{13}CS} lines, respectively, will help break this degeneracy. 
The CS is ringed and the \ce{H_2CS} follows this same spatial distribution with a \ce{H_2CS}/\ce{CS} ratio of 0.22$\pm$0.04 at 30~K and 0.18$\pm$0.04 at 60~K (assuming N(\ce{CS}) =  N(\ce{C^{34}S})$\times$22). This value is consistent with what was observed for disks by \citet{2019ApJ...876...72L} who report that the \ce{H_2CS}/\ce{CS} ratio in disks is more similar to pre-stellar cores than protostellar envelopes and photo-dissociation regions where the \ce{H_2CS}/\ce{CS} values are higher. We also detect SO which lies interior to the CS emission. This change in volatile S carrier across the disk is likely tracing the change in C/O at the \ce{H_2O} snowline. From Figure~3 the N(CS)/N(SO) ratio is $1.0\pm$0.2 at 10~au and then increases radially to $>10$ by 100~au. Since the SO data are convolved to a lower spatial resolution (0.7") the minimum N(CS)/N(SO) is likely lower than 1.0 since the emission is more compact at 0.2" and the CS already shows a gas cavity in the 0.7" data.  
Compared to other Herbig disks, this ratio sits in the middle with Oph-IRS~48 having a disk averaged N(CS)/N(SO) $<$0.01 and HD~142527 a disk averaged ratio $>$1 \citep{2021A&A...651L...6B, 2023arXiv230406382T}.
An elevated C/O in the outer disk gas where the CS is present is also supported by the rings of \ce{HC_3N} and \ce{c-C_3H_2}. \citet{2023NatAs...7...49C} show with chemical models that in a C/O$>1$ UV-driven chemistry leads to the efficient formation of \ce{HC_3N} in the gas-phase. 

\begin{figure*}
    \centering
    \includegraphics[width=0.95\hsize]{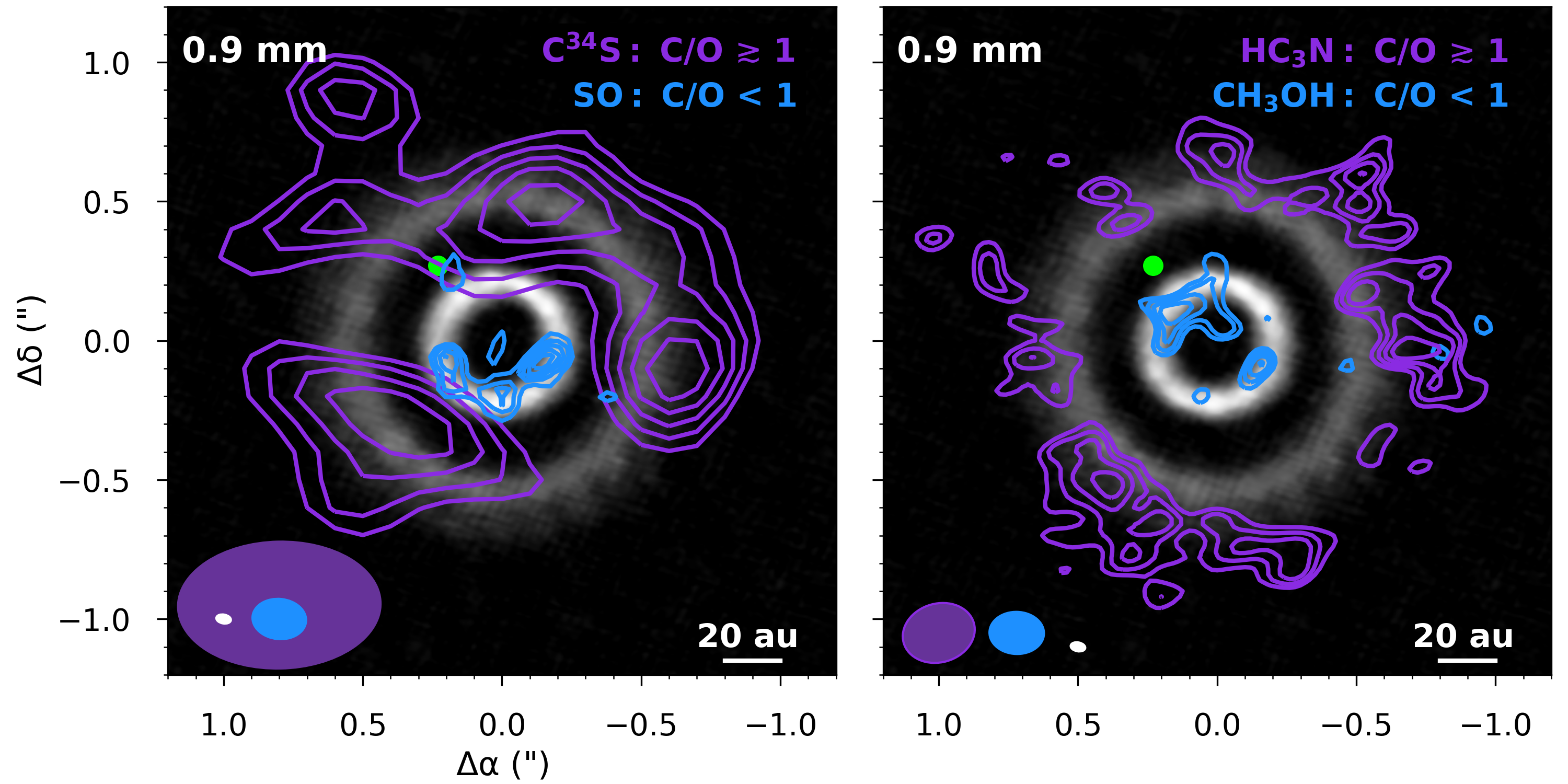}
    \caption{HD~169142 0.9~mm continuum emission (grey-scale) with coloured contours showing the 60, 70, 80 and 90\% integrated flux values for \ce{C^{34}S} and \ce{SO} in the left panel and \ce{HC_3N} and \ce{CH_3OH} in the right panel. The green dot highlights the location of the giant planet at ~37~au \citep{2019A&A...623A.140G, 2022MNRAS.517.5942G, 2023arXiv230211302H}.
    The different beams for each molecular line are shown in the same colours as the contours.}
    \label{fig:3}
\end{figure*}

\subsection{Organic content}

We detect a number of organic molecules in the HD~169142 disk: \ce{H_2CO}, \ce{CH_3OH}, \ce{HC_3N} and \ce{c-C_3H_3}. It is interesting to compare the distribution of \ce{H_2CO} and \ce{CH_3OH} as they both share a formation route on the grains \citep{2009A&A...505..629F}. 
HD~169142 hosts a warm disk with $\mathrm{T_{dust}}>20$~K and therefore limited CO freeze-out. This implies the inheritance of \ce{CH_3OH} ices rather than in-situ formation \citep{2021NatAs...5..684B}. 
The disk temperature also favours the formation of the observed \ce{H_2CO} in the gas-phase rather than on the grains \citep{2021ApJ...906..111T}. We find that the two strongest \ce{H_2CO} lines we detect have central cavities with a drop in emission where the \ce{CH_3OH} is detected. This difference may be a result of excitation since these lines have upper energy levels of 21 and 35~K whereas we expect the \ce{CH_3OH} to be present at temperatures $>$100~K. To potentially support this, the third \ce{H_2CO} line we (weakly) detect has an upper energy level of 68~K and shows compact emission like the \ce{CH_3OH} (see Figure D1). 
The difference between \ce{H_2CO} and \ce{CH_3OH} can also be due to the gas-phase formation of the \ce{H_2CO} in the outer disk which does not occur for \ce{CH_3OH}. 
In Figure~3 we show the radial ratio of N(\ce{CH_3OH})/N(\ce{H_2CO}). Within the \ce{H_2O} snowline line at 10~au this ratio is 140$\pm$40 which decreases sharply to $<$3 beyond $\approx$60~au. This is similar to the trend shown for the HD~100546 disk \citep{2021NatAs...5..684B} although, the peak of the ratio is lower in HD~100546. A N(\ce{CH_3OH})/N(\ce{H_2CO}) ratio $>$1 is also consistent with thermal ice desorption observed in the Class II disk Oph-IRS~48, observations of Class I protostars, and cometary ice abundances \citep{2016A&A...595A.117J,2017RSPTA.37560252B,2021A&A...651L...5V}. 
In the colder outer disk, the data are not sensitive enough to detect the same non-thermally desorbed reservoir of \ce{CH_3OH} as seen for HD~100546 and TW~Hya where N(\ce{CH_3OH})/N(\ce{H_2CO})$\approx$1-2 \citep{2019A&A...623A.124C,2021NatAs...5..684B}. 
The \ce{H_2CO} extends out to 200~au where a component of the \ce{H_2CO} could originate from non-thermal desorption of \ce{H_2CO} rich ices at the edge of the millimetre dust disk. 

The \ce{H_2CO}, \ce{DCO^+} and \ce{DCN} line fluxes for HD~169142 are consistent with the trends derived in \citet{2023arXiv230302167P}, where the Herbig disks have lower line fluxes relative to \ce{C^{18}O} (a proxy for total gas mass) than the T-Tauri disks indicating less active cold chemistry (see Figure~4). 
We take the HD169142 \ce{C^{18}O} $J=2-1$ line flux of 3.9~Jy~km~s$^{-1}$ from \citet{2017A&A...600A..72F}. 
The similarities in the \ce{DCO^+}, \ce{DCN} and \ce{H_2CO} in the three Herbig disks may indicate that CO freeze-out is not significantly impacting the primary formation/destruction routes of these species in Herbig disks. 

At the edge of mm-dust disk is also where the rings of \ce{HC_3N} and \ce{c-C_3H_2} are peaking. \citet{2021ApJS..257....9I} report clear rings of \ce{HC_3N} and \ce{c-C_3H_2} (and \ce{CH_3CN}) in the AS~209, HD~163296 and MWC~480 protoplanetary disks. In comparison to HD~169142, the \ce{HC_3N} and \ce{c-C_3H_2} emission rings from these sources, all fall well within the radial extent of the mm-dust disks. However, there are some similarities between these disks. For the HD~163296 disk the peak of the \ce{H_2CO} emission is just beyond the peaks of the \ce{HC_3N} and \ce{c-C_3H_2} rings \citep{2021ApJS..257....3L, 2021ApJS..257....6G, 2021ApJS..257....9I} and for HD~169142 we find that the \ce{H_2CO} is more radially extended in than the \ce{HC_3N} and \ce{c-C_3H_2}. Higher angular resolution \ce{H_2CO} observations are required to properly determine the relationship between the dust, \ce{H_2CO} and the large organics \ce{CH_3OH}, \ce{HC_3N} and \ce{c-C_3H_2}.

\subsection{Connections to planet composition}

Multiple complementary studies indicate that there are at least two giant planets in the HD~169142 system: one within the central mm-dust cavity (likely a massive companion rather than a planet) and another located in the gap between mm-dust rings at 37~au - HD~169142~b \citep{2019A&A...623A.140G, 2022A&A...663A..23L,  2022MNRAS.517.5942G, 2023arXiv230211302H}. Using the molecular line observations of the HD~169142 disk we can infer the chemical make-up of the gas in the proximity of these planets. 
Figure~5 shows mm-dust emission, the expected location of HD~169142~b and contours tracing the emission from \ce{^{34}CS}, \ce{SO}, \ce{CH_3OH} and \ce{HC_3N}. Under the assumption that \ce{SO} and \ce{CH_3OH} trace gas with C/O$<$1 any planet forming within the central dust cavity will be forming within the \ce{H_2O} snowline and therefore, able accrete gas with a bulk C/O$<$1. In contrast, HD~169142~b is forming beyond the water snowline and given the temperature structure of the disk (See Figure E1), beyond the \ce{CO_2} snowline too. 
Since the CO snowline lies well beyond the dust disk, HD~169142~b should be located within the CO snowline. From this, the most simple interpretation is that HD~169142~b is accreting gas with an elemental C/O ratio of 1.0 \citep{2011ApJ...743L..16O}. From Figure~3 the N(CS)/N(SO) ratio at the location of HD~169142~b is 2.0$\pm$0.4. Comparing to astrochemical models, this is broadly consistent with our interpretation of an overall C/O$\approx$1 in this region of 
the disk \citep{2018A&A...617A..28S, 2020A&A...638A.110F}.
{The [CI] also traces the gas in the outer disk and has been detected in two other Herbig disks - HD~163296 and HD~142527 \citep{2022ApJ...941L..24A, 2023arXiv230406382T}. Chemical models show that [CI] is expected to trace the upper layers of the disk and can be used to determine the carbon abundance in the disk atmosphere \citep{2016A&A...592A..83K}. Further work to specifically model [CI] in disks will allow for both comparisons of disk C/H and C/O to giant exoplanet atmospheres.} The gas composition in the vicinity of the planet may be altered if the planet is locally heating the disk, leading to a C/O$<1$ in the gas as suggested for HD~100546 \citep{2023A&A...669A..53B} and tentatively traced by the SO in HD~169142 \citep[see Figure 5;][]{Law2023}.

\section{Conclusion}
We have analysed ALMA data of the HD~169142 disk and presented observations of 13 different species. Our main conclusions are as follows:
\begin{itemize}
\item We present the first detections of \ce{N_2H^+}, \ce{CH_3OH}, [CI], \ce{DCN}, \ce{CS}, \ce{C^{34}S}, \ce{^{13}CS}, \ce{H_2CS}, \ce{H_2CO}, \ce{HC_3N} and \ce{c-C_3H_2} in the HD~169142 disk. 

\item The line ratios and abundance ratios of \ce{DCO^+}, DCN, \ce{H_2CO}, CS and \ce{H_2CS} detected in HD~169142 are consistent with observations of the other Herbig disks HD~163296 and MWC~480. 

\item The \ce{N_2H^+} line flux from the HD~169142 disk is low relative to HD~163296 and MWC~480. This comparison and the temperature structure in the disk model from \citet{2018A&A...614A.106C} indicates the lack of CO freeze-out in the HD~169142 disk but, if there is a CO freeze-out in this disk it is likely beyond the mm-dust disk edge ($\approx$100~au) and out at $\approx$150~au. 

\item The detection of \ce{CH_3OH} and \ce{SO} from the inner disk can be explained by the sublimation of ices putting the \ce{H_2O} snowline at the edge of the inner mm-dust $\approx$20~au. This is consistent with the brightness temperature of CO exceeding 100~K in this region of the disk \citep{2022A&A...663A..23L}. 

\item HD~169142 is now the third warm Herbig disk where the signature of \ce{CH_3OH} ice sublimation has been detected. Since \ce{CH_3OH} needs to form in the cold phase, this further supports the hypothesis that complex ices can survive the disk formation process. 
\item The variation in the radial column density ratio of CS to SO supports gas-phase C/O variations in the HD~169142 disk across the \ce{H_2O} snowline.
The large organic molecules \ce{HC_3N} and \ce{c-C_3H_2} may be tracing gas with a C/O$\approx$1 and appear in rings at the edge of the mm-dust disk.


\item We estimate that the giant planet HD~169142~b is forming between the \ce{CO} and \ce{H_2O} snowlines and given the temperature structure of the disks likely between the \ce{CO_2} and \ce{CO} snowlines. This is a region of the disk where C/O$\approx$1.0. However, this ratio may be lowered if the planet is heating the disk locally resulting in the sublimation of O-rich ices. Future higher angular resolution observations will be able to further characterise the gas composition in the vicinity of this planet.  If there is another giant planet in the central cavity $<$20~au this is within the \ce{H_2O} snowline and therefore the planet will be accreting gas with C/O$<$1. 
\end{itemize}

Many of the line detections presented in this work are serendipitous detections in continuum spectral windows in archival ALMA data. As continuum observations with ALMA reach higher sensitivities the likelihood of picking up such lines increases. We urge the community to be mindful of the frequency ranges and spectral resolution (FDM mode) used for continuum observations so that if lines are detected there is the best opportunity possible to use them for science. 

\begin{acknowledgements}
    We thank Karin Öberg and Ewine van Dishoeck for their useful comments on this paper. This work makes use of the following ALMA data:  2012.1.00799.S, 2015.1.00806.S, 2016.1.00344.S, 2016.1.00346.S, 2018.1.01237. ALMA is a partnership of ESO (representing its member states), NSF (USA) and NINS (Japan), together with NRC (Canada), MOST and ASIAA (Taiwan), and KASI (Republic of Korea), in cooperation with the Republic of Chile. The Joint ALMA Observatory is operated by ESO, AUI/NRAO and NAOJ. We acknowledge assistance from Allegro, the European ALMA Regional Centre node in the Netherlands. Astrochemistry in Leiden is supported by funding from the European Research Council (ERC) under the European Union’s Horizon 2020 research and innovation programme (grant agreement No. 101019751 MOLDISK). This work has used the following additional software packages that have not been referred to in the main text: Astropy, IPython, Jupyter, Matplotlib and NumPy \citep{Astropy,IPython,Jupyter,Matplotlib,NumPy}.
\end{acknowledgements}

\bibliographystyle{aa} 
\bibliography{aanda_revised} 

\newpage
\onecolumn

\begin{appendix}

\section{ALMA Observations}

\begin{table}[h!]
    \caption{Summary of ALMA observations used in this work.}
    \centering
    \begin{tabular}{c c c cc  c c c c }
    \hline 
      Project ID   &PI &  Band & Date & No. Ant. & Baselines & Integration Time & M.R.S.  \\
     & & & && [m] & [min] & ["]  \\ \hline \hline
     2012.1.00799.S & Honda, M. &7 & 2015-07-26 & 41 & 15–1574 & 41.4 & 1.3 \\ 
     && 7 & 2015-07-27 & 41 & 15–1574 & 21.2 & 1.4  \\ 
     && 7 & 2015-08-08 & 43 & 35–1574 & 41.4 & 1.3  \\ 
     2015.1.00806.S & Pineda, J. & 7 & 2015-12-06 & 32 & 19–7716 & 25.3 & 0.4  \\ 
    2016.1.00344.S& P\'erez, S.  &6 & 2016-10-04 & 40  & 19-3144 & 33.3 & 8.5  \\ 
                  & & &  2017-07-05 & 44  & 21-2647  & 33.2 & 7.7 \\ 

    2016.1.00346.S & Tsukagoshi, T. & 8&  2017-04-05 & 40 & 15-279 & 2.92 & 5.9  \\ 
    2018.1.01237.S & Mac\'ias, E. & 7 & 2019-04-10 & 43 & 15-500 & 81.0 & 5.7  \\  
    \hline
    \end{tabular}
    \label{tab:my_label}
\end{table}

\section{Comparisons of matched filter and \textit{GoFish} detections of \ce{N_2H^+}}

\begin{figure}[h!]
    \centering
    \includegraphics[width=0.95\hsize]{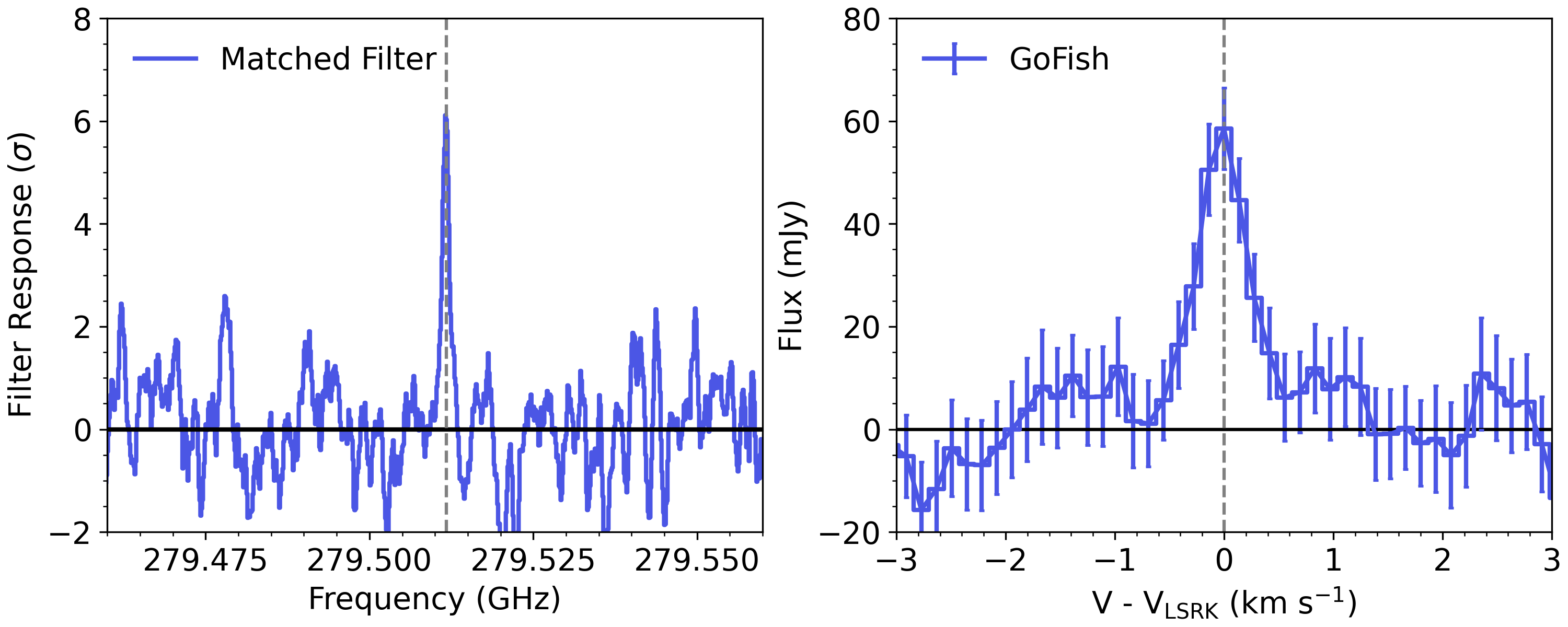}
    \caption{Matched filter response (left) and \textit{GoFish} spectra (right) for \ce{N_2H+} in the HD~169142 disk \citep{2018AJ....155..182L,GoFish}. The matched filter response is for a 0-200~au Keplerian mask.}
    \label{fig:my_label}
\end{figure}

\newpage
\section{Spectra extracted from continuum spectral windows}

\begin{figure}[h!]
    \centering
    \includegraphics[width=\hsize]{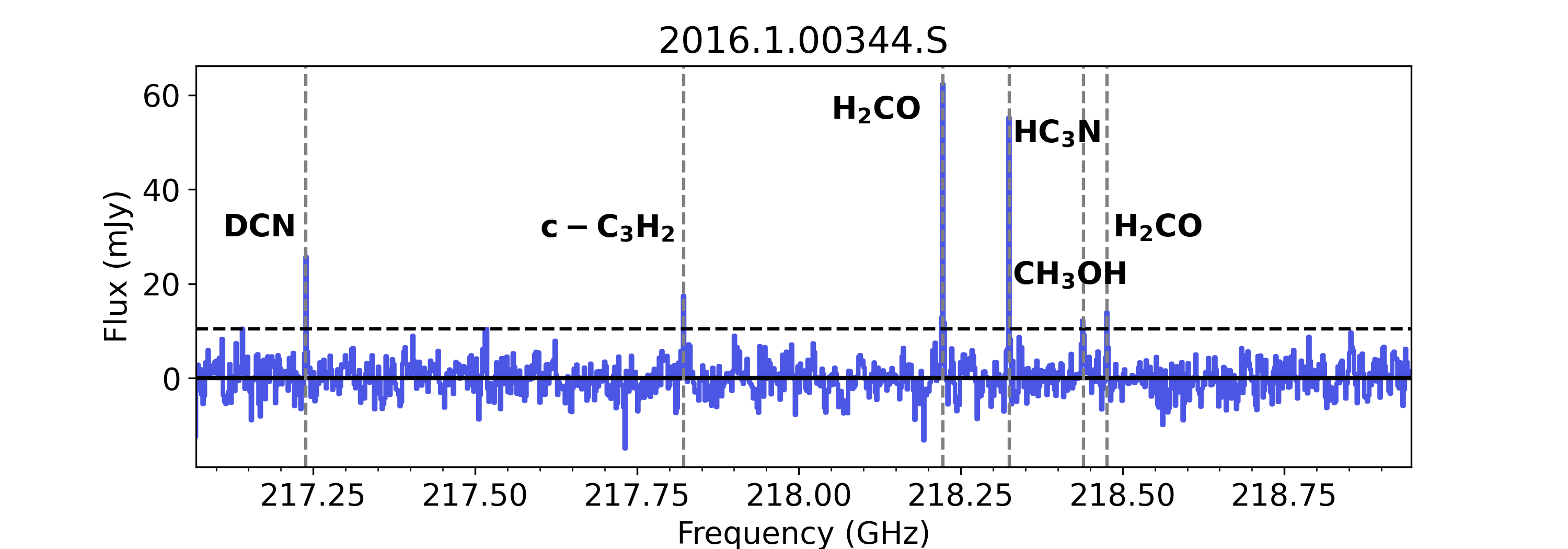}
    \includegraphics[width=\hsize]{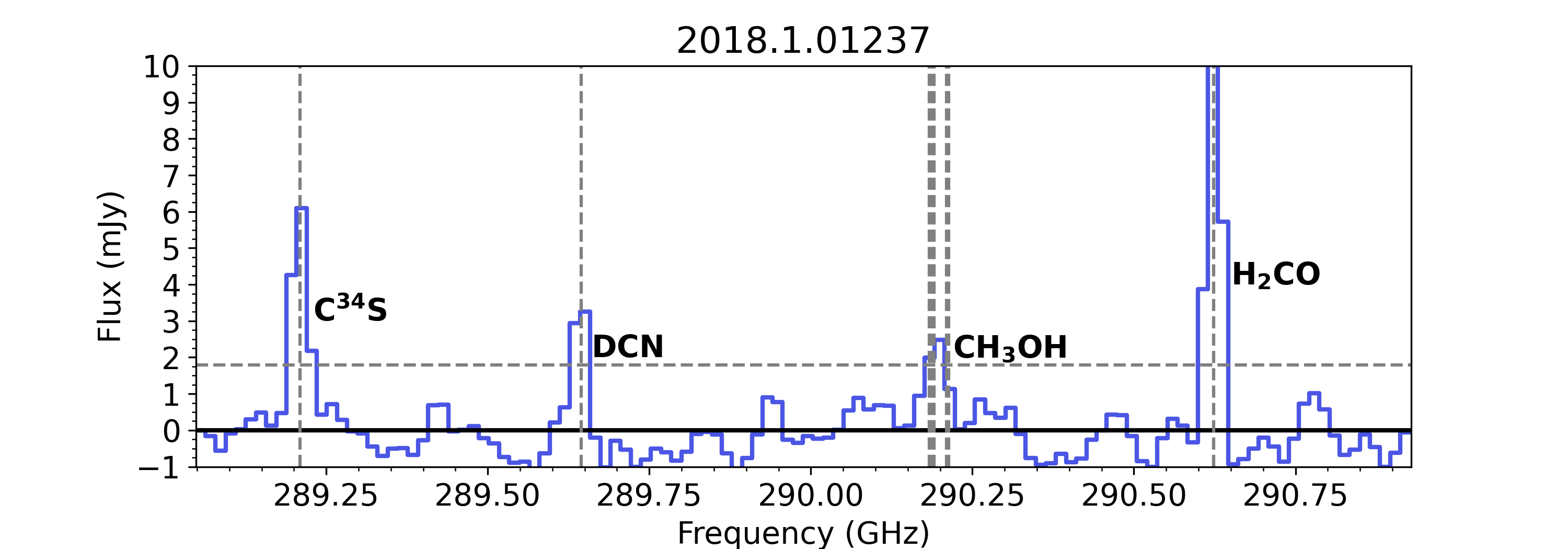}
    \caption{Spectra extracted from the continuum spectral windows using a 1\farcs0 circular aperture for the two data sets noted in the plot titles. The molecules detected are highlighted with vertical lines at their respective frequencies as listed in Table~1. The dashed horizontal line marks the 3~$\sigma$ level where $\sigma$ is calculated as the rms in the line free channels of the spectrum. We note that to encompass the total flux of most of the lines a 2\farcs0 aperture is required but the 1\farcs0 aperture shows both the detection of compact \ce{CH_3OH} and the more extended molecules, e.g., \ce{H_2CO}.} 
    \label{fig:my_label}
\end{figure}

\newpage
\section{Supplementary line images}
\begin{figure}[h!]
\centering
    \includegraphics[width=0.65\hsize]{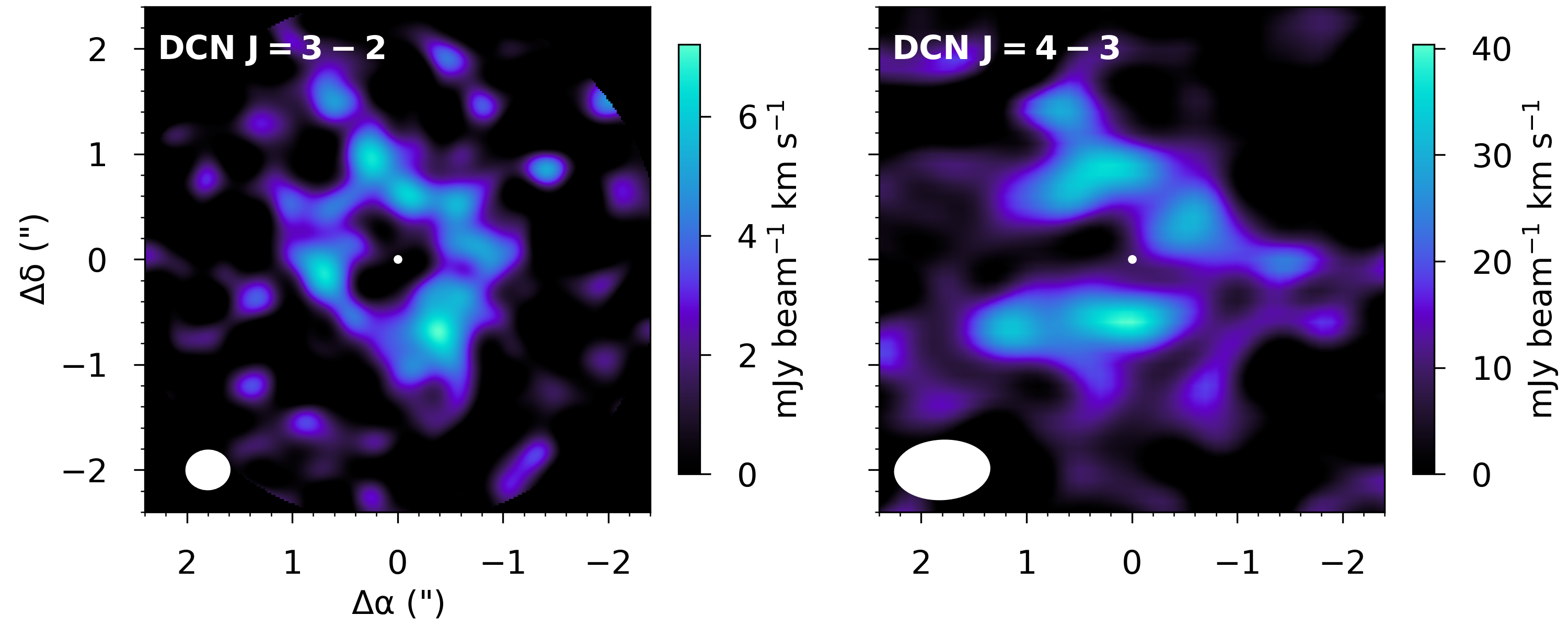} \\
    \includegraphics[width=0.67\hsize]{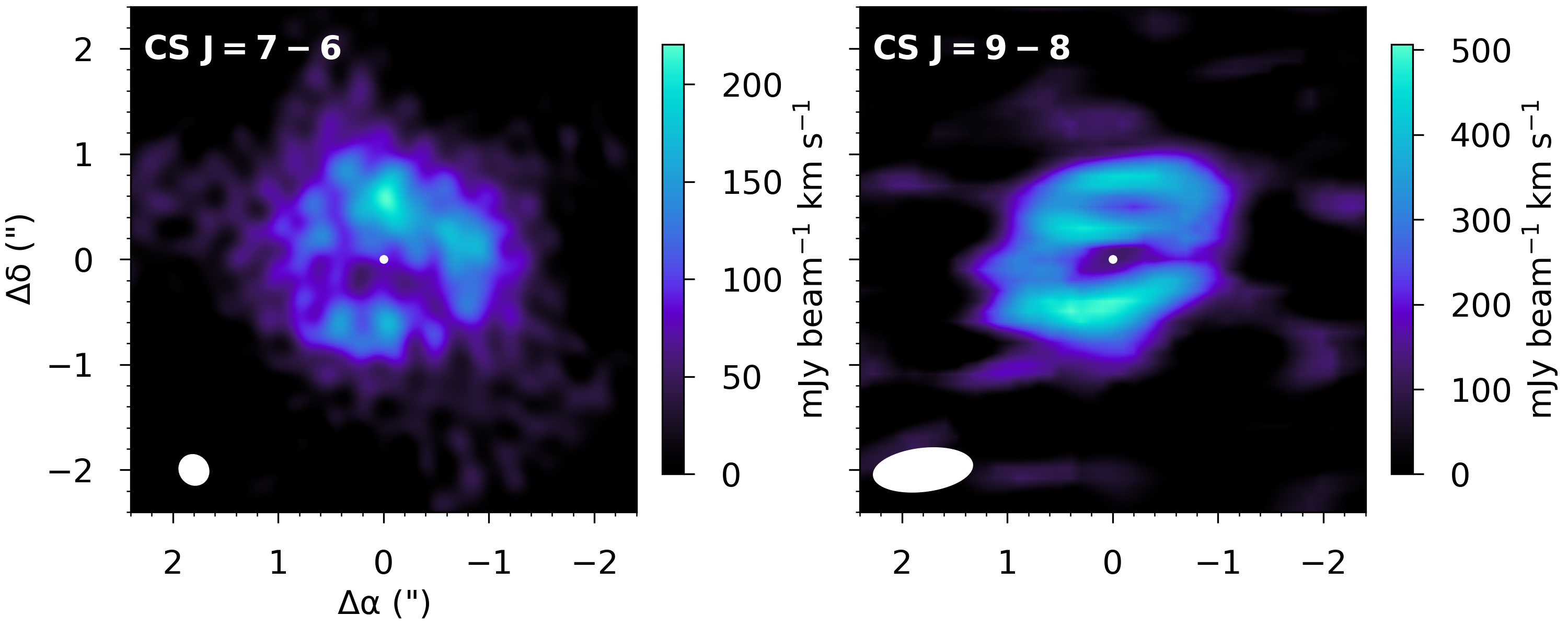} \\
     \includegraphics[width=\hsize]{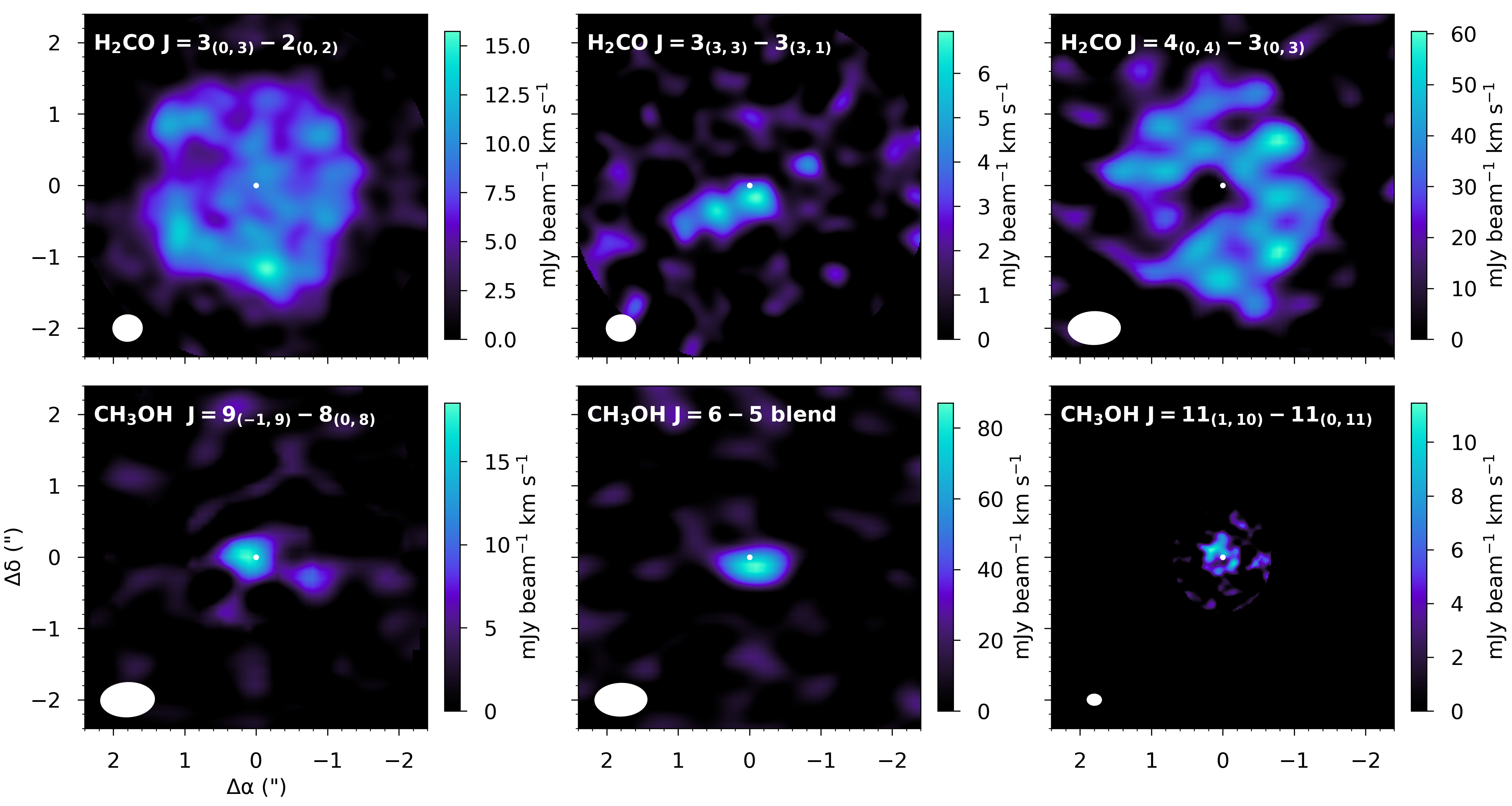}    
     \caption{Integrated intensity maps of additional molecular transitions detected in the HD~169142 disk.}
    \label{fig:my_label}
\end{figure}

\newpage
\section{HD~169142 model dust temperature}

\begin{figure}[h!]
    \centering
    \includegraphics[trim={0cm 0cm 0cm 0cm},clip,width=0.5\hsize]{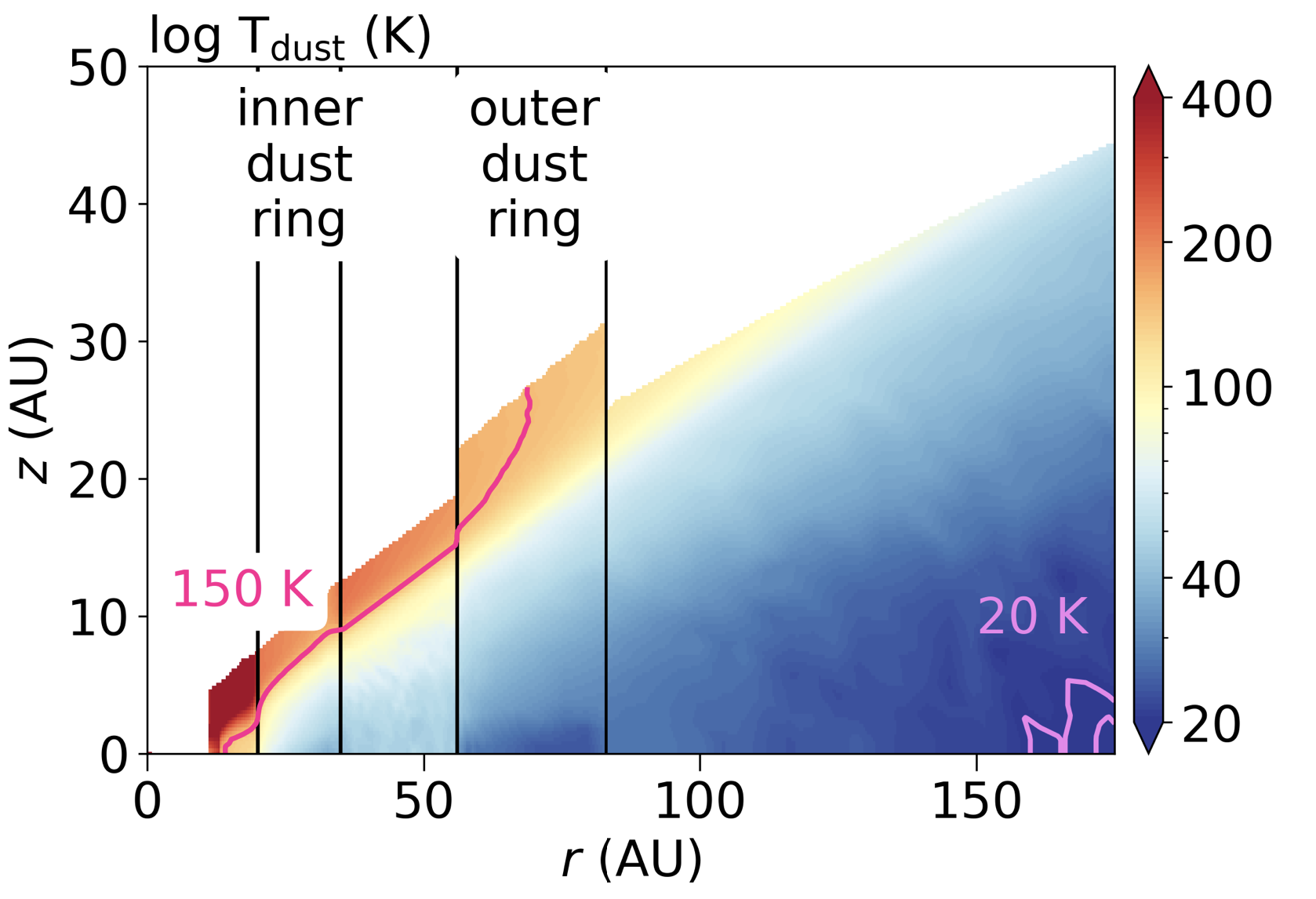}
    \caption{Dust temperature in the fiducial DALI model from \citet{2018A&A...614A.106C}.}
    \label{fig:my_label}
\end{figure}

\end{appendix}
\end{document}